\documentclass[a4paper, 10pt, conference]{ieeeconf}      

\IEEEoverridecommandlockouts                              

\usepackage{amsmath} 
\usepackage{amssymb}  
\newtheorem{lemma}{Lemma}

\title{\LARGE \bf
Fast symmetric matrix inversion using modified Gaussian elimination
}

\author{Anton Kochnev, Nikolai Savelov
\thanks{Anton Kochnev is currently a student of Platov South-Russian State Polytechnic University (NPI), Russia.
	{\tt\small avkochnev@yandex.ru}}%
\thanks{Nikolai Savelov is a doctor of technical science, professor of Platov South-Russian State Polytechnic University (NPI), Russia.
	{\tt\small savelovn@mail.ru}}%
}

\begin{document}

\maketitle
\thispagestyle{empty}
\pagestyle{empty}

\begin{abstract}

In this paper we present two different variants of method for symmetric matrix inversion, based on modified Gaussian elimination. Both methods avoid computation of square roots and have a reduced machine time's spending. Further, both of them can be used efficiently not only for positive (semi-) definite, but for any non-singular symmetric matrix inversion. We use simulation to verify results, which represented in this paper.

\end{abstract}

\section{INTRODUCTION}

Symmetric matrix inversion is one of the most important problem for many practical tasks e.g. analysis of electrical circuits with inductance elements \cite{Chua}, synthesis of Kalman or Wiener filters \cite{Happonen}, using of finite element method \cite{Streng}.\par
Existing symmetric matrix inversion methods are Cholesky decomposition, LDL decomposition \cite{Golub}, bordering method \cite{Krylov}, and the most efficient Krishnamoorthy\,-\,Menon's method (based on Cholesky decomposition, and requires $\displaystyle \frac{n^3}{2}+\frac{n^2}{2}$ operations with $n$ square roots computation) \cite{Krishnamoorthy,Poznan}.\par
The aim of this paper is to propose a symmetric matrix inversion method, which reduces machine time spending compared to Krishnamoorthy\,-\,Menon's method by avoiding of square root computations. Moreover, this fact allow us to use the proposed method for efficient inverse of symmetric matrix not only with strict diagonal dominance, but without diagonal dominance as well.

\section{MODIFIED GAUSSIAN ELIMINATION}


In this section shows modified Gaussian elimination, which proposed method based on \cite{Savelov85}-\cite{Savelov_Kochnev}.\par
Let there is a system of linear equations (\ref{SLE}),
\begin{equation}
\label{SLE}
Ax=b,
\end{equation}
where $A\in{\mathbb{C}}^{{n}\times{n}}$, $x\in{\mathbb{C}}^{{n}\times{1}}$, $b\in{\mathbb{C}}^{{n}\times{1}}$. During modified Gaussian elimination an addition matrix $F:F\in{\mathbb{C}}^{{n}\times{n}}$ changes instead of matrix $A$, but addition memory is not necessary in this case \cite{Savelov_Kochnev}.\par
Let vector $x$ consist of two types of variables: required, which should be find during elimination, and unrequired, which is not interesting for researcher. Re-solution of (\ref{SLE}) after some changes of $A$ could be done with reduced number of multiplications and divisions, using formulae from \cite{Savelov85}-\cite{Savelov_Kochnev}.\par
Let $a_i$  be an  $i^{th}$ column of $A$; let $F^m$ be a matrix $F$ after $m^{th}$ change; let $f^m_i$ be  an $i^{th}$ row of $F^m$, $m= \overline{0,n}$; let $F^0=I$, where $I$ is an identity matrix. Then solution of (\ref{SLE}) could be done with formulae (\ref{Modification}) below:

\begin{itemize}
	\item $f^{m+1}_i=f^m_i$ for $i<m+1$, $m=\overline{1,n-1}$, and $x_i$ is an unrequired variable;
	\item$\displaystyle f^{m+1}_{m+1}=\frac{1}{f^m_{m+1}a_{m+1}}f^{m}_{m+1}$ for $m=\overline{0,n-1}$.
	If $f^m_{m+1}a_{m+1}=0$, then two rows of $F$ should be permitted. Such permutation is always possible for non-singular matrix $A$ (see \cite{Savelov85}-\cite{Savelov_Kochnev});
	\item $f^{m+1}_i=f^m_i-(f^m_i a_{m+1}) f^{m+1}_{m+1}$ for other required $i$;
	\item $x_i=f^n_i b$ for any required $x_i$.
\end{itemize}
\begin{equation}
\label{Modification}
\end{equation}

It can easily be checked that $f^n_i a_i=1$, $f^n_i a_j=0$, for any required $x_i$ and for $j=\overline{1,n}$, $j\ne i$. Further, if all elements of $x$ are required variables, then $F^n=A^{-1}$ . Matrix inversion using modified Gaussian elimination requires $n^3$ multiplications and divisions. If only the last element $x_n$ of $x$ is a required variable, then Gaussian elimination requires $\displaystyle \frac{n^3}{3}+\frac{n^2}{2}+\frac{n}{6}$ multiplies and divisions. Let $p$ be a number of required variables in $x$: $x_i, i=\overline{n-p+1,n}$ is a required variable. Then number of multiplies and divisions for Gaussian elimination could be determined with formula $\displaystyle \frac{n^3}{3}+\frac{n^2}{2}+\frac{n}{6}+p^2n-pn-\frac{p^3}{3}+\frac{p^2}{2}-\frac{p}{6}$ (see \cite{Savelov_Kochnev}).


\section{PROPOSED METHOD}

In this section shows a method for efficient symmetric matrix inversion based on modified Gaussian elimination.

\subsection{First variant of the proposed method}

The first variant of the proposed method consist of two different stages. On the first stage we use formulae (2) for $p=1$, and only $x_n$ is a required variable (indeed, it is  a valid proposition for $p=0$ as well). On the second stage we use addition formulae described below.\par

Let us introduce some notation.\par

Let $f^{m+1}_{i j}$ be an element from $i^{th}$ row and $j^{th}$ column of $F^{m+1}$; let $a_{i j}$ be an element from $i^{th}$ row and $j^{th}$ column of $A$; let $S^{m+1}_f$be a submatrix of $F^{m+1}$, such that \mbox{$S^{m+1}_f=(f^{m+1}_{i j})$}, $i=\overline{1,m+1}$, $j=\overline{1,m+1}$; let $S_a$ be a submatrix of $A$, such that $S_a=(a_{i j})$, $i=\overline{1,m+1}$, $j=\overline{1,m+1}$.\par

It is easily shown that after the first stage $F$ is a lower triangular matrix. After the second stage $F$ became a matrix $A^{-1}$, as if we use (2) for $p=n$.\par

For symmetric matrix inversion using modified Gaussian elimination that is enough to use only lower triangular matrix. To prove this statement, we need a lemma \ref{lemma_1}.\par

\begin{lemma}
	\label{lemma_1}
	Let $F^{m+1}$ is a matrix from (\ref{Modification}) for \mbox{$p=n$}, \mbox{$m=\overline{0,n-1}$}. Then $S^{m+1}_f=(S_a)^{-1}$.
\end{lemma}

\proof

Using (\ref{Modification}), let us consider 4 cases. \par
Case 1. Let $i=m+1$. Then
\begin{align}
	&f^{m+1}_{i} a_i=f^{m+1}_{m+1} a_{m+1},\notag\\
	&f^{m+1}_{m+1} a_{m+1}=\frac{1}{f^m_{m+1}a_{m+1}}f^m_{m+1} a_{m+1}=1,\notag\\
	&m=\overline{0,n-1};\notag
\end{align}

Case 2. Let $j=m+1$. Then
\begin{align}
	&f^{m+1}_i a_j=f^{m+1}_i a_{m+1},\notag\\
	&f^{m+1}_i a_{m+1}=f^m_i a_{m+1}-f^m_i a_{m+1} f^{m+1}_{m+1} a_{m+1}=0,\notag\\
	&m=\overline{0,n-1}, i=\overline{1,m}; \notag
\end{align}

Case 3. Let $i=m$. Then
\begin{align}
	&f^{m+1}_i a_i=f^{m+1}_m a_m,\notag\\
	&f^{m+1}_m a_m=f^m_m a_m-f^m_m a_{m+1} f^{m+1}_{m+1} a_m,\notag\\
	&f^{m+1}_m a_m=f^m_m a_m-\frac{f^m_m a_{m+1}}{f^{m}_{m+1}a_{m+1}}f^{m}_{m+1} a_m=1;\notag
\end{align}
It can be checked easily for $\forall i:i<m+1$, $m=\overline{0,n-1}$;

Case 4. Let $j=m$. Then
\begin{align}
	&f^{m+1}_i a_j=f^m_i a_m-f^m_i a_{m+1} f^{m+1}_{m+1} a_m=0;\notag
\end{align}
It can be checked easily for $\forall j$: $j<m+1$, $j\ne i$, \mbox{$m=\overline{0,n-1}$}.

\endproof

Since lemma \ref{lemma_1}, it follows that $S^{m+1}_f$ is a symmetric matrix.\par
This statement is needed for the second stage of the method. Let $F^k$ be a matrix $F$ after $k^{th}$ change, $k=\overline{0,n}$, where $F^0$ is a lower triangular matrix after the first stage, and $F^n=A^{-1}$. Let $f^k_{i j}$ is an element from $i^{th}$ row and $j^{th}$ column of $F^k$. Then the second stage describes with formulae below:\par

\begin{itemize}
	\item $f^{k+1}_{i j}=f^k_{i j}$ for $i=\overline{k+1,n}$, $j=\overline{1,n}$, $k=\overline{0,n-1}$, and for $i=\overline{1,k}$, $j>i$, $k=\overline{1,n-1}$;
	\item $f^{k+1}_{i j}=f^{k}_{i j}+\displaystyle \frac{f^{k+1}_{k+1 i} f^{k+1}_{k+1 j}}{f^{k+1}_{k+1 k+1}}$ for $i=\overline{1,k}$, $j=\overline{1,i}$, \mbox{$k=\overline{1,n-1}$}.
\end{itemize}
\begin{equation}
\label{Second_stage}
\end{equation}

It is easily to prove that $A^{-1}=F^n+{(F^n-D)^T}$, where $D$ is a diagonal matrix, such that $d_{i i}=f^n_{i i}$, $i=\overline{1,n}$.\par

Number of multiplications and divisions for the first stage of the method describes with formula $\displaystyle \frac{n^3}{3}+\frac{n^2}{2}+\frac{n}{6}$, as we note earlier.\par

It is easily shown that number of multiplications and divisions for the second stage describes with formula \mbox{$\displaystyle \frac{n^3}{6}+\frac{n^2}{2}-\frac{2n}{3}$}.\par

Both stages for symmetric matrix inversion using the first variant of the proposed method requires $\displaystyle \frac{n^3}{2}+{n^2}-\frac{n}{2}$ multiplications and divisions. It is less then requirements of Cholesky decomposition or LDL decomposition (see table 1).\par

Let us remark that proposed method avoid square root computations; this considerably reduce machine time spending, and make it possible to use proposed method not only for positive determined, but for any invertable symmetric matrices as well.\par

\subsection{Second variant of the proposed method}

The second variant of the proposed method consist of only one stage.\par

Suppose, that

\begin{equation}
\label{Second_variant}
F^{m+1}=F^m_c+\Delta F^m,
\end{equation}

where $F^m_c=F^m$, but all elements from $(m+1)^{th}$ row of $F^m_c$ are zeros.\par

Let $f^m_{i j}$ be an element of $i^{th}$ row and $j^{th}$ column of $F^m$; let $f^m_{\bullet j}$ be a $j^{th}$ column of $F^m$.\par

For explanation of the following formulae, we need a lemma \ref{lemma_2}.\par

\begin{lemma}
	\label{lemma_2}
	Let $F^{m+1}$ is a matrix from (\ref{Modification}) for $p=n$, $m=\overline{0,n-1}$. Then $\Delta F^m$ from (\ref{Second_variant}) be $\Delta F^m=(f^{m+1}_{\bullet m+1}f^m_{m+1})$.
\end{lemma}

\proof
From (\ref{Second_variant}) $\Delta F^m=F^{m+1}-F^m_c$. \par
Using (\ref{Modification}), we get \par
1.
\begin{align}
	&f^{m+1}_{m+1}=\frac{1}{f^m_{m+1}a_{m+1}}f^{m}_{m+1},\notag\\
	&f^{m+1}_{m+1}=f^{m+1}_{m+1 m+1}f^{m}_{m+1},\notag\\
	&f^{m+1}_{m+1 m+1}f^{m}_{m+1}=f^{m+1}_{m+1}-0, m=\overline{0,n-1};\notag
\end{align}

2.
\begin{align}
	&f^{m+1}_i=f^m_i-(f^m_i a_{m+1}) f^{m+1}_{m+1},\notag\\
	&f^{m+1}_i=f^m_i-\frac{f^m_i a_{m+1}}{f^m_{m+1} a_{m+1}} f^{m}_{m+1},\notag\\
	&f^{m+1}_i=f^m_i-({f^m_i a_{m+1}}) f^{m}_{m+1 m+1} f^{m}_{m+1},\notag\\
	&f^{m+1}_i=f^m_i+f^{m+1}_{i m+1}f^m_{m+1},\notag\\
	&f^{m+1}_{i m+1}f^m_{m+1}=f^{m+1}_i-f^m_i,\notag\\
	&m=\overline{0,n-1}, i=\overline{1,n}, i\ne{m+1}.\notag
\end{align}

\endproof

If we combine lemma \ref{lemma_1} with lemma \ref{lemma_2}, we get formulae below:

\begin{itemize}
	\item $\displaystyle f^{m+1}_{i m+1}=\frac{1}{f^m_{m+1} a_{m+1}}f^m_{m+1 i}$ for $m=\overline{0,n-1}$, \mbox{$i=\overline{1,m+1}$},
	\item $f^{m+1}_{i m+1}=-(f^m_ia_{m+1})f^{m+1}_{m+1 m+1}$ for $m=\overline{0,n-2}$, \mbox{$i=\overline{m+2,n}$},
	\item $f^{m+1}_{i j}=f^m_{i j}+f^{m+1}_{i m+1}f^m_{m+1 j}$ for $m=\overline{1,n-1}$, $j=\overline{1,m}$, $i=\overline{j,n}$, $i\ne m+1$,
	\item $f^{m+1}_{m+1 j}=f^{m+1}_{j m+1}$ for $m=\overline{1,n-1}$, $j=\overline{1,m}$,
	\item $f^{m+1}_{i j}=f^{m+1}_{i j}$ for $m=\overline{2,n-1}$, $j=\overline{2,m}$, \mbox{$i=\overline{1,j-1}$},
	\item $f^{m+1}_{i j}=f^m_{i j}$ for $m=\overline{0,n-2}$, $i=\overline{1,n}$, $j=\overline{m+2,n}$.
\end{itemize}
\begin{equation}
\label{second_a}
\end{equation}

The formulae (\ref{second_a}) describe an idea of the method in detail, but for practical tasks it is better to use different formulae:

\begin{itemize}
	\item $\displaystyle f^{m+1}_{m+1 j}=\frac{1}{f^m_{m+1} a_{m+1}}f^m_{m+1 j}$ for $m=\overline{0,n-1}$, \mbox{$j=\overline{1,m+1}$},
	\item $f^{m+1}_{i m+1}=-(f^m_ia_{m+1})f^{m+1}_{m+1 m+1}$ for $m=\overline{0,n-2}$, \mbox{$i=\overline{m+2,n}$},
	\item $f^{m+1}_{i j}=f^m_{i j}+f^{m+1}_{i m+1}f^m_{m+1 j}$ for $m=\overline{1,n-2}$, $i=\overline{m+2,n}$, \mbox{$j=\overline{1,m}$},
	\item $f^{m+1}_{i j}=f^{m}_{i j}+f^{m+1}_{m+1 i}f^m_{m+1 j}$ for $m=\overline{1,n-1}$, $i=\overline{1,m}$, $j=\overline{1,i}$,
	\item $f^{m+1}_{i j}=f^{m+1}_{i j}$ for $m=\overline{0,n-2}$, $i=\overline{1,n}$, \mbox{$j=\overline{m+2,n}$},
	\item $f^{m+1}_{i j}=f^m_{i j}$ for $m=\overline{1,n-1}$, $i=\overline{1,m}$, \mbox{$j=\overline{i+1,m+1}$}.
\end{itemize}
\begin{equation}
\label{second_b}
\end{equation}

It can easily be shown that (\ref{second_a}) and (\ref{second_b}) are equivalent.\par

Number of multiplications and divisions for the second variant of the proposed method describes with formulae \mbox{$\displaystyle\frac{n^3}{2}+\frac{n^2}{2}$}. It is less then requirements of Cholesky decomposition, LDL decomposition or Krishnamoorthy\,-\,Mennon method. It is the same requirements as for bordering method, but it should be noted that  bordering method could not be use for inversion of matrix with $M_{i i}\ne 0$, $i=\overline{1,n}$, where $M_{i j}$ is a minor of $A$, $i=\overline{1,n}$, $j=\overline{1,n}$.\par

At the same time, proposed method could be use for inversion of matrix with $M_{i i}=0$, if $A$ is not a singular matrix.\par

Let us remark that the second variant of the proposed method avoid square root computations as well.

\section{SIMULATION SETUP}

In order to demonstrate advantages of the proposed algorithms, we use MATLAB based simulation via different CPUs. We give results for Intel Core i5-3230M 2.60 GHz below (Intel Pentium Dual Core T 2390 1.86 GHz and Intel Atom N450 1.67 GHz gives familiar results). We compare proposed algorithms with the most efficient notable algorithms for symmetric matrix inversion: Cholesky decomposition \cite{Samarskij}, LDL decomposition \cite{Golub}, and Krishnamoorthy\,-\,Mennon method \cite{Krishnamoorthy}, \cite{Poznan}, \cite{Matlab}. We generate table with full equations, which describes number of multiplications, divisions and square roots computation for every noted method.\par

The first row of the table 1 describes Cholesky decomposition and solving of systems of linear equations $L B = I$, and $L^T A^{-1} = B$. The second row describes LDL decomposition and solving of SLE $\tilde{L} X = I$, $D \tilde{B} = X$, and $\tilde{L^T} A^{-1} = \tilde{B}$. The third row describes matrix inversion using Krishnamoorthy\,-\,Mennon method, based on Cholesky decomposition \cite{Krishnamoorthy}, \cite{Poznan}. The fourth row describes the first variant, and the fifth row describes the second variant of the proposed method.\par

Experiment 1. Inversion of a real symmetric matrix with strict diagonal dominance. Let $q_{theor}$ be a number of multiplications and divisions, and $s_{theor}$ be a number of square root computations, determined with formulae from the table 1. Let $q_{pract}$ and $s_{pract}$ be numbers of operations, determined with counter variables from MATLAB scripts. Results of the experiment are given in tables 2 and 3.\par

Experiment 2. Inversion of real symmetric matrices of order $n$ with strict diagonal dominance.
Let $t$ be a time for matrix inversion; let $norm$ be a second $norm:norm=\lVert A^{-1}_m-A^{-1}_{inv}\rVert _2$, where $A^{-1}_m$ is a matrix, inverted via one of described methods, and $A^{-1}_{inv}$ is a matrix, inverted via MATLAB function $inv(A)$. Results of the experiment are given in tables 4 and 5.\par

Experiment 3. Inversion of real symmetric matrices of order $n$ without diagonal dominance. Results of the experiment are given in the table 6.\par

\section{SIMULATION RESULTS}

From tables 2 and 3 we can conclude that formulae from the table 1 are correct.\par
From tables 4 and 5 we can conclude that both variants of the proposed method provide notable reduction of machine time spending and has a good accuracy.\par
From the table 6 we can conclude that both variants of the proposed method increase advantages for matrices without diagonal dominance. Let us remark that it is especially important for inductance matrix inversion.\par

\section{CONCLUSIONS}

We propose a new method for symmetric matrix inversion based on modified Gaussian elimination with avoiding of square root computations. Proposed method could be useful for any scientific and technical problem with symmetric matrix inversion, especially if matrix has not a diagonal dominance.

\clearpage
\begin{table}[h]
	\begin{center}
		\begin{tabular}{|p{2.5cm}|p{2.5cm} |p{2.5cm}|}
			\hline METHOD OF MATRIX INVERSION & NUMBER OF MULTIPLICATIONS AND DIVISIONS & NUMBER OF SQUARE ROOT COMPUTATIONS \\
			\hline Cholesky decomposition & $\displaystyle \frac{n^3}{2}+\frac{3n^2}{2}$ & $\displaystyle n$ \\ 
			\hline  LDL decomposition & $\displaystyle \frac{2n^3}{3}+\frac{n^2}{2}-\frac{n}{6}$ & 0 \\[2mm] 
			\hline  Krishnamoorthy - Mennon's method (based on Cholesky decomposition) & $\displaystyle \frac{n^3}{2}+\frac{n^2}{2}$ & $\displaystyle n$ \\ 
			\hline  The first variant of the proposed method & $\displaystyle \frac{n^3}{2}+{n^2}-\frac{n}{2}$ & 0 \\ 
			\hline  The second variant of the proposed method & $\displaystyle \frac{n^3}{2}+\frac{n^2}{2}$ & 0 \\ 
			\hline 
		\end{tabular}
	\end{center}
	Table 1. Table summarizing the number of operations for inversion of a matrix with strict diagonal dominance via different methods.
\end{table}

\begin{table}[h]
	\begin{center}
		\begin{tabular}{|p{2.5cm}|p{1.0cm}|p{1.0cm}|p{1.0cm}|p{1.0cm}|}
			\hline METHOD OF MATRIX INVERSION & $q_{theor}$ & $q_{practr}$ & $s_{theor}$ & $s_{pract}$ \\ 
			\hline Cholesky decomposition & 515000 & 515000 & 100 & 100 \\ 
			\hline LDL decomposition & 671650 & 671650 & 0 & 0 \\ 
			\hline Krishnamoorthy - Mennon's method (based on Cholesky decomposition) & 505000 & 505000 & 100 & 100 \\ 
			\hline The first variant of the proposed method & 509950 & 509950 & 0 & 0 \\ 
			\hline The second variant of the proposed method & 505000 & 505000 & 0 & 0 \\ 
			\hline 
		\end{tabular}
	\end{center}
	Table 2. Table summarizing the number of operations for inversion of a matrix with strict diagonal dominance via different methods ($n=100$).
\end{table}

\begin{table}[h]
	\begin{center}
		\begin{tabular}{|p{2.5cm}|p{1.0cm}|p{1.0cm}|p{1.0cm}|p{1.0cm}|}
			\hline METHOD OF MATRIX INVERSION & $q_{theor}$ & $q_{practr}$ & $s_{theor}$ & $s_{pract}$ \\ 
			\hline Cholesky decomposition & 62875000 & 62875000 & 500 & 500 \\ 
			\hline LDL decomposition & 83458250 & 83458250 & 0 & 0 \\ 
			\hline Krishnamoorthy - Mennon's method (based on Cholesky decomposition) & 62625000 & 62625000 & 500 & 500 \\ 
			\hline The first variant of the proposed method & 62749750 & 62749750 & 0 & 0 \\ 
			\hline The second variant of the proposed method & 62625000 & 62625000 & 0 & 0 \\ 
			\hline 
		\end{tabular}
	\end{center}
	Table 3. Table summarizing the number of operations for inversion of a symmetric matrix with strict diagonal dominance via different methods ($n=500$).
\end{table}

\begin{table}[h]
	\begin{center}
		\begin{tabular}{|p{2.5cm}|p{1.0cm}|p{1.0cm}|p{1.0cm}|p{1.0cm}|}
			\hline METHOD OF MATRIX INVERSION & $n=100$ & $n=300$ & $n=500$ & $n=1000$ \\ 
			\hline Based on Cholesky decomposition & 0.384 s. & 9.499 s. & 41.96 s. & 329.7 s. \\ 
			\hline LDL decomposition & 0.332 s. & 7.950 s. & 33.23 s. & 261.0 s. \\ 
			\hline Krishnamoorthy - Mennon's method (program by A. Krishnamoorthy \cite{Matlab}) & 0.125 s. & 2.950 s. & 12.12 s. & 96.84 s. \\ 
			\hline Krishnamoorthy - Mennon's method (program with element-by-element access) & 0.113 s. & 2.793 s. & 12.44 s. & 103.3 s. \\ 
			\hline The first variant of the proposed method & 0.046 s. & 0.743 s. & 3.086 s. & 27.13 s. \\ 
			\hline The second variant of the proposed method & 0.046 s. & 0.718 s. & 2.845 s. & 25.64 s. \\ 
			\hline 
		\end{tabular} 
	\end{center}
	Table 4. Table summarizing times of numerical computations for inversion of a symmetric matrix with strict diagonal dominance via different methods.
\end{table}

\begin{table}[h]
	\begin{center}
		\begin{tabular}{|p{2.5cm}|p{1.0cm}|p{1.0cm}|p{1.0cm}|p{1.0cm}|}
			\hline METHOD OF MATRIX INVERSION & $n=100$ & $n=300$ & $n=500$ & $n=1000$ \\ 
			\hline Based on Cholesky decomposition & 9.4E-19 & 1.3E-18 & 1.9E-18 & 3.0E-18 \\ 
			\hline LDL decomposition & 1.0E-18 & 1.5E-18 & 2.0E-18 & 3.0E-18 \\ 
			\hline Krishnamoorthy - Mennon's method (program by A. Krishnamoorthy \cite{Matlab}) & 9.4E-19 & 1.3E-18 & 1.9E-18 & 3.0E-18 \\ 
			\hline Krishnamoorthy - Mennon's method (program with element-by-element access) & 9.4E-19 & 1.3E-18 & 1.9E-18 & 3.0E-18 \\ 
			\hline The first variant of the proposed method & 1.5E-18 & 2.5E-18 & 4.1E-18 & 6.7E-18 \\ 
			\hline The second variant of the proposed method & 1.5E-18 & 2.5E-18 & 4.1E-18 & 6.7E-18 \\ 
			\hline 
		\end{tabular} 
	\end{center}
	Table 5. Table summarizing ${\lVert A^{-1}_m-A^{-1}_{inv}\rVert}_2$ of numerical computations for inversion of a symmetric matrix with strict diagonal dominance via different methods.
\end{table}

\begin{table}[h]
	\begin{center}
		\begin{tabular}{|p{2.5cm}|p{1.0cm}|p{1.0cm}|p{1.0cm}|p{1.0cm}|}
			\hline METHOD OF MATRIX INVERSION & $n=100$ & $n=300$ & $n=500$ & $n=1000$ \\ 
			\hline Based on Cholesky decomposition & 0.448 s. & 11.16 s. & 50.47 s. & 401.2 s. \\ 
			\hline LDL decomposition & 0.328 s. & 7.420 s. & 33.22 s. & 264.4 s. \\ 
			\hline Krishnamoorthy - Mennon's method (program by A. Krishnamoorthy \cite{Matlab}) & 0.153 s. & 6.185 s. & 54.42 s. & 861.7 s. \\ 
			\hline Krishnamoorthy - Mennon's method (program with element-by-element access) & 0.318 s. & 10.66 s. & 73.34 s. & 1021 s. \\ 
			\hline The first variant of the proposed method & 0.045 s. & 0.742 s. & 3.093 s. & 27.05 s. \\ 
			\hline The second variant of the proposed method & 0.045 s. & 0.701 s. & 2.836 s. & 25.30 s. \\ 
			\hline 
		\end{tabular} 
	\end{center}
	Table 6. Table summarizing times for inversion of a symmetric matrix without diagonal dominance via different methods.
\end{table}

\end{document}